# Transverse orbital angular momentum and polarization entangled spatiotemporal structured light

Hsiao-Chih Huang[1,2], Kefu Mu[2], Hui Min Leung[2*], and Chen-Ting Liao[1*]

[1]Department of Physics, Indiana University, Bloomington, Indiana 47405, U.S.A.

[2]Department of Intelligent Systems Engineering, Indiana University, Bloomington, Indiana 47405, U.S.A

Intra-system entanglement occurs between non-separable modes within the same system. For optical systems, the various degrees of freedom of light represent different modes, and the potential use of light to create higher dimensional classical entangle states offers a promising potential to drive new technological developments. In this work, we present experimental results demonstrating the orthogonality between transverse orbital angular momentum (t-OAM) of different spatiotemporal topological charges, a previously unverified property of t-OAM. Based on those results, we developed methods to create and characterize a novel family of t-OAM and polarization entangled spatiotemporal structured light. We further provide theoretical analysis to support our study of the entanglement between those modes. By demonstrating the feasibility of leveraging t-OAM as a new family of modes for classical entanglement, our work represents a new advancement towards higher dimensional classical entanglement strategies.

## 1. Introduction.

Entanglement is a fundamental feature utilized in quantum information science and technology. In this domain, information is encoded non-locally in a joint state of a system, enabling functionalities that surpass those in classical systems. Entanglement underpins transformative applications, such as quantum cryptography and quantum teleportation. Two-partite entangled biphoton pairs, generated through spontaneous parametric down-conversion, serve as foundational building blocks widely utilized in photonic quantum computing and quantum networking [1], [2]. Related to entanglement is the concept of non-separability. From a mathematical perspective, a function is considered non-separable if it cannot be factorized into two or more independent terms. For instance, for analytical functions $f$ and $g$, the function $F_{sep} = f_1g_1 + f_1g_2 = f_1(g_1 + g_2)$ is separable while the function $F_{nonsep} = f_1g_1 + f_2g_2$ is not [3], [4], [5]. Importantly, the functions $f_1$ and $f_2$, along with $g_1$ and $g_2$, must constitute an orthogonal basis. When applied to the concept of entanglement, $f_i$ and $g_i$ represent different degrees of freedom (DoFs), often referred to as *modes*, with their subscripts indicating their dimensionalities. Quantum entangled states arise from inter-system non-separability across one or more DoFs between two or more particles. Those that involve two or more DoFs are known as multipartite entanglement or high-dimensional entanglement.

Over the past three decades, it has become evident that intra-system entanglement within a single particle can also manifest in the classical domain through the presence of non-separable modes. This is commonly referred to as classical entanglement or mode entanglement [6], [7], [8]. Unlike quantum entanglement, mode entanglement in the classical domain lacks multiple-particle nonlocality, making it physically impossible to achieve non-separability between distinct systems. Instead, the non-separability in this context refers to the correlations between the system's distinguishable modes, such as path, spin, momentum, and energy. For optical systems, the various DoFs of light, such as wavelength, polarization, and angular momentum of light represent different modes. Consider a simple example of classical entanglement between conventional, *longitudinal* orbital angular momentum (OAM) and linear polarization, represented as $f$ and $g$, respectively. The complex optical field as a result of their mode entanglement can be expressed as $U = f_1g_1 + f_2g_2$. Using $l$ to represent the conventional topological charge of the OAM modes (i.e., the first DoF, A) and either H or V to represent horizontal or vertical polarizations (i.e., the second DoF, B), we can then adapt Dirac's notation to express the total state vector of such a mode-entangled vector vortex beam as $|\Psi\rangle \propto |l_1\rangle_A|V\rangle_B + |l_2\rangle_A|H\rangle_B$. The study of the analogous relationship between the quantum and classical cases is facilitated by their similar mathematical expressions.

## 2. Transverse OAM as a new family of modes.

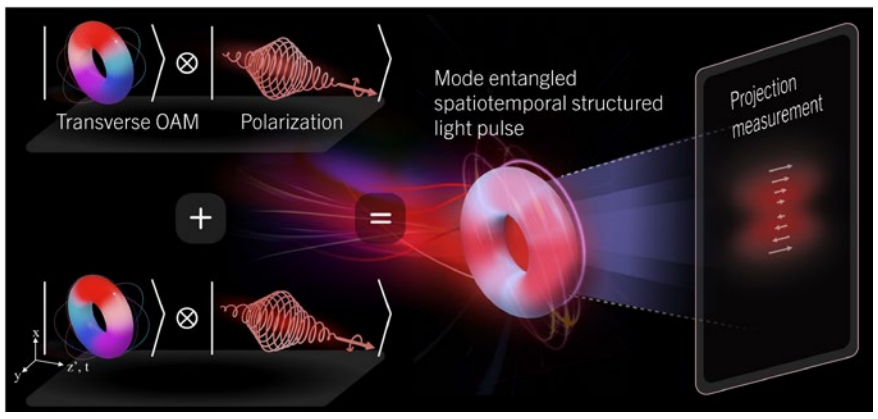

**Figure 1.** Schematic illustration of a new family of mode-entangled state of light (center), constructed by spatiotemporal structured vector light fields. This new state is formed by combining two pulses, one with positive t-OAM and right-hand circular polarization (top left), and the other with the opposite and orthogonal t-OAM and left-hand circular polarization (bottom left). White arrows (right) indicate polarization directions.

Longitudinal OAM of light, first identified over three decades ago [9], is characterized by a time-averaged OAM vector that is parallel to the beam's propagation direction. In contrast, the time-averaged OAM vector of light carrying *transverse* OAM (t-OAM) is perpendicular to the propagation direction. Due to its space-time dependent optical characteristics, t-OAM is also referred to as spatiotemporal OAM of light [10], [11]. This phenomenon was first explored theoretically in 2005 [12]. Experimentally, t-OAM was observed in laser filaments and termed spatiotemporal optical vortices (STOVs) [13]. The successful experimental generation of STOVs using controllable optical setups sparked substantial research interest [14], [15]. Studies on STOVs have rapidly expanded, addressing topics such as their diffraction behavior [16], characterization techniques [17], [18], nonlinear conversions [19], [20], and applications in communication and quantum information processing [21], [22]. A particularly noteworthy application is the recent experimental demonstration of sixteen STOV strings for image transmission [23], which opens up new opportunities for communication and quantum information processing. Higher dimensional quantum and classical entanglement offers a promising avenue for fundamental research and has the potential to drive new technological advancements in entanglement-based applications. Consequently, the combination, conversion, and creation of different DoFs represent powerful techniques that are of significant research interests [24]. Nevertheless, the use of t-OAM as a DoF in entanglement has not been previously explored in either the quantum or classical domain.

*Corresponding authors: Chen-Ting Liao (liao3@iu.edu); Hui Min Leung (huileung@iu.edu)

A scalar, linearly polarized t-OAM light beam can be represented as $E(x,y,z,t) \propto U e^{iq\Phi_{ST}(x,t)} e^{i(k_z z-\omega t)} = U e^{iq\Phi_{ST}} e^{ik_z(z-v_g t)}$, where $U = U(x,y,z,t)$ is the complex amplitude of the pulsed laser, which includes trivial Gaussian envelopes in the x-axis, $e^{-x^2/x_0^2}$, y-axis, $e^{-y^2/y_0^2}$, and time-axis $e^{-t^2/t_0^2}$, where $x_0$, $y_0$ and $t_0$ are nominal pulse widths in space and time, such as the full width at half maximum (FWHM) beam sizes and pulse duration, respectively. We can also define z'=z-$v_g$t, the space-like coordinate in the moving reference frame of a laser pulse travelling with group velocity $v_g$, and $z' = z - v_g t_0$. In our experiments, $v_g$ is close to the vacuum speed of light (i.e., $v_g \sim c$) given that our experiments are performed in air, a dilute and linear media, with low peak power from the laser. Here we adapt near-paraxial condition and quasi-monochromatic condition for our experimental and theoretical considerations. The quasi-monochromatic condition is justified given our many-cycled femtosecond pulsed beam centered at 1030nm wavelength with 6nm FWHM bandwidth, where fractional bandwidth is small (Δλ/λ = 6/1030 = 0.0058 << 1). Assuming a pure t-OAM as those from line-STOVs [13], the *azimuthal* phase dependence in space-time is expressed as $\Phi_{ST} = \Phi_{ST}\left(x, t = (z-z')/v_g\right) = tan^{-1}(\frac{x/x_0}{t/t_0})$. We can then rewrite the expression of t-OAM carrying light pulse as $E(x,z') \propto U(x,z')\left(\frac{z'}{\eta z_0'} \pm i\frac{x}{x_0}\right)^{|q|} e^{iq\Phi_{ST}(x,t)}$, where $\eta$ is the asymmetry parameter that accounts for specific experimental conditions that generate t-OAM of light, assuming insignificant dispersion, divergence, and diffraction. For simplicity, we consider the ideal case where $\eta = 1$. The t-OAM states are quantified by the mode index $q$, also known as the spatiotemporal topological charge, where q=0, ±1, ±2, etc.

In order to construct t-OAM and polarization mode entangled spatiotemporal structured light, we coherently combined two pulsed beams, each possessing two DoFs with states orthogonal to each other. Figure 1 shows a schematic that illustrates the framework for creating this new family of mode-entangled light states. In this example, one of the beams (top left) would carry t-OAM of q=+1 (as indicated by the hue with 2$\pi$ phase winding in the space and time) and R polarization (spiral electric fields in red). The other beam (bottom left) would carry t-OAM of q=-1 (as indicated by the hue with -2$\pi$ phase winding that goes in the opposite in direction) and L polarization. The coherently combined beam result in a mode entangled vector pulsed beam. Its projection leads to a position-dependent intensity (depicted in red) and spatial- and temporal-dependent polarization (depicted as white arrows) profiles. Adapting Dirac's notation, we can then rewrite the total state vector of such a mode entangled, spatiotemporal structured vector beam as

$$|\Psi(x,y,z,t)\rangle \propto |q=+1\rangle_A \otimes |R\rangle_B + e^{i\delta}|q=-1\rangle_A \otimes |L\rangle_B = |{+1}\rangle_A|R\rangle_B + e^{i\delta}|{-1}\rangle_A|L\rangle_B, \quad (1)$$

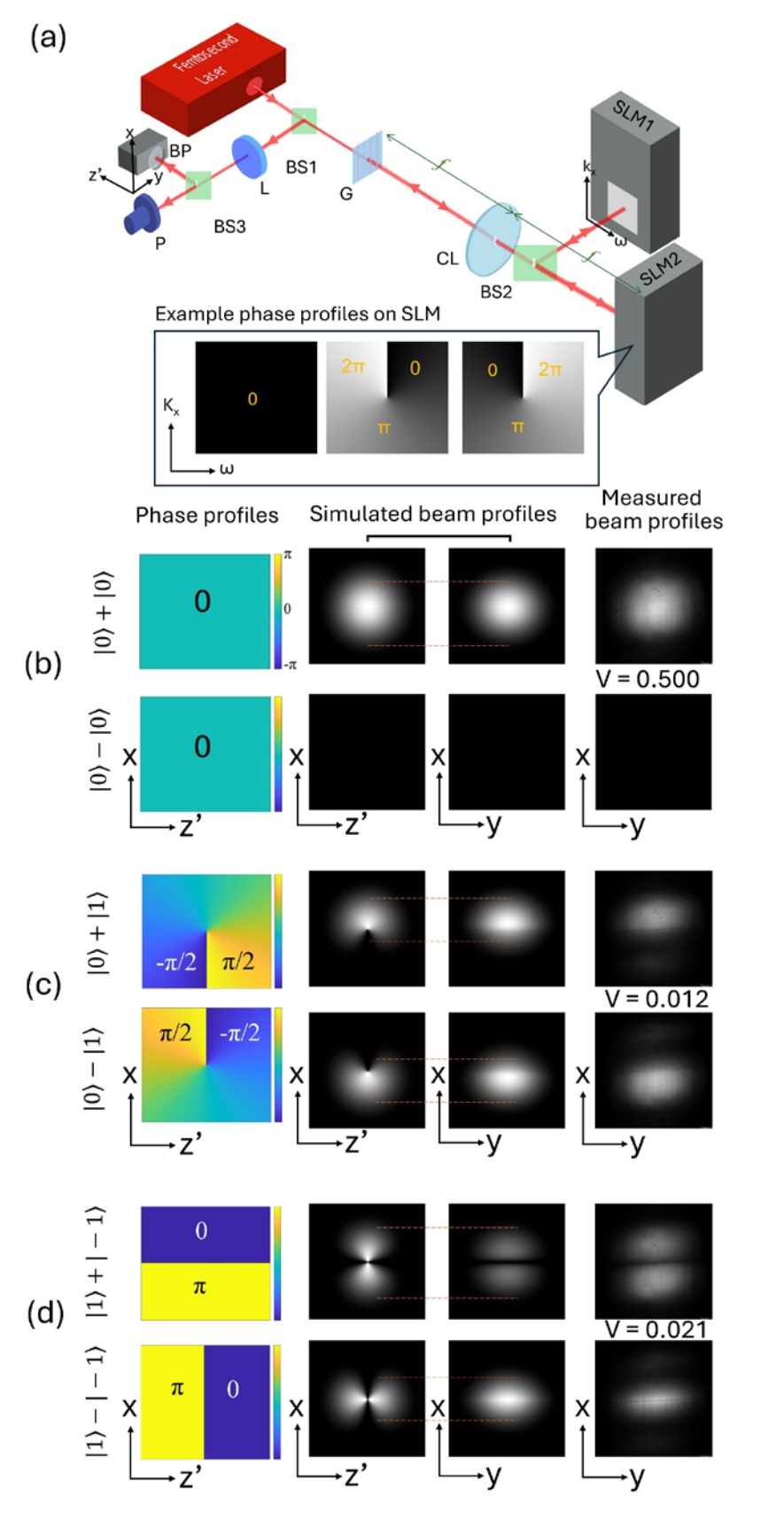

**Figure 2.** Phase/intensity profiles and orthogonality between light carrying different t-OAM charges. (a) A schematic setup used for generating light carrying t-OAM and establishing the orthogonality between them. The setup consists of a Michelson-like interferometer within a pulse shaper. A beam profiler captures the intensity profile of the superposition of light with two different t-OAM charges at the x-y plane. As a measure of their orthogonality and interferability, the visibility, V, of their superposition was determined. BP: Beam profiler; BS1, BS2, BS3: Plate beamsplitters; P: Power meter; L: Spherical lens; CL: Cylindrical lens; G: Grating; SLM1, SLM2: Spatial light modulators. (b), (c), (d) Results of |q=0⟩±|q=0⟩, |q=0⟩±|q=1⟩, and |q=1⟩±|q=-1⟩, respectively. The first column depicts the simulated phase profiles of the in-phase and out-of-phase superpositions of two beams. The second and third columns show the corresponding simulated intensity profiles at the x-z' and x-y planes, respectively. The dotted red lines between the two highlights the fact that the latter is the projection of the former. The last column shows measured beam intensity profile in the x-y plane.

where $e^{i\delta}$ is the relative phase between the two beams. The two DoFs, denoted by subscripts A and B, are t-OAM and circular polarizations of light, respectively. When $e^{i\delta} =$ $e^{i(0)} = e^{i(2\pi)} = 1$, the first and the second term are in-phase; when $e^{i\delta} = e^{i(\pi)} = -1$, they are out-of-phase. Although we demonstrated such mode entanglement using only two lowest charges of t-OAM (i.e., q=+1 and -1), we emphasize that the t-OAM index could theoretically be infinite and the generation of q=100 has recently been demonstrated [25].

In order to use t-OAM as a new DoF to construct a mode entangled light, as expressed in Equation (1), it is essential to verify that t-OAM states of different charges are orthogonal to one other. In the following Section 3, through experimental verification and theoretical analysis, we present results demonstrating the orthogonality between t-OAM of different spatiotemporal topological charges, a previously unverified property of t-OAM. After establishing the orthogonality of t-OAM states and, thus, satisfying the prerequisite condition for mode entanglement, we proceeded to experimentally construct a new family of t-OAM and circular polarization mode entangled vector beams of light. We also introduced methods to characterize the created mode entangled vector beams, which consists of imaging spectrometry and projection measurements with interferometry. Lastly, in section 5, we conclude our work with future directions and potential new applications based on this newly established classical entangled light.

## 3. Orthogonal tests of t-OAM states of light.

Our experimental setup for generating t-OAM states of light and testing their orthogonality is shown in Fig. 2(a). A 200-fs pulsed Gaussian laser beam centered at 1030 nm wavelength (OneFive Origami), 6nm FWHM bandwidth, is sent into a custom 4f pulse shaper that consists of a cylindrical lens, CL (f=30cm), and a transmission grating, G (1600 lines/mm). Unlike other common pulse shapers used to generate STOVs, our experimental setup included a Michelson-like interferometer that comprised a 50:50 thin plate beamsplitter (BS2) and two spatial light modulators (SLMs). It acts as a beam amplitude division within the pulse shaper, generating two identical beams that were sent to two spatial light modulators, SLM1 and SLM2. The SLMs were mounted on translational stages to enable experimentally adjustable optical delays. Light of different topological charges was generated by imparting the desired Laguerre-Gaussian (LG) phase profiles via the SLMs onto the beams. As indicated on Fig. 2(a), the SLM coincides with the $k_x$-ω Fourier plane (or x-ω plane) of the pulse shaper. The beams that returned from SLM1 and SLM2 overlapped at BS2, forming a Michelson-like interferometer with tunable relative delays (i.e., temporal phases). The superimposed, coherently combined two beams were then sent through a spherical lens (L) and into either a 2D beam profiler (indicated as BP in Fig. 2(a)) or a power meter (indicated as P in Fig. 2(a)). The orientation of the x-y-z' axes is shown in Fig 2(a), and it can be seen that the 2D plane of the beam profiler lies in the x-y plane of the SLMs in the pulse shaper. S-polarized light is consistently used for all experiments described in Section 3. A more detailed experimental setup can be found in the Supplemental Materials.

The orthogonality test results consisting of simulated and experimental results of the superpositions of two beams, where the two beams represent either |q=0>±|q=0>, |q=0>±|q=1>, or |q=1>±|q=-1> are shown in Fig. 2(b), (c), and (d),

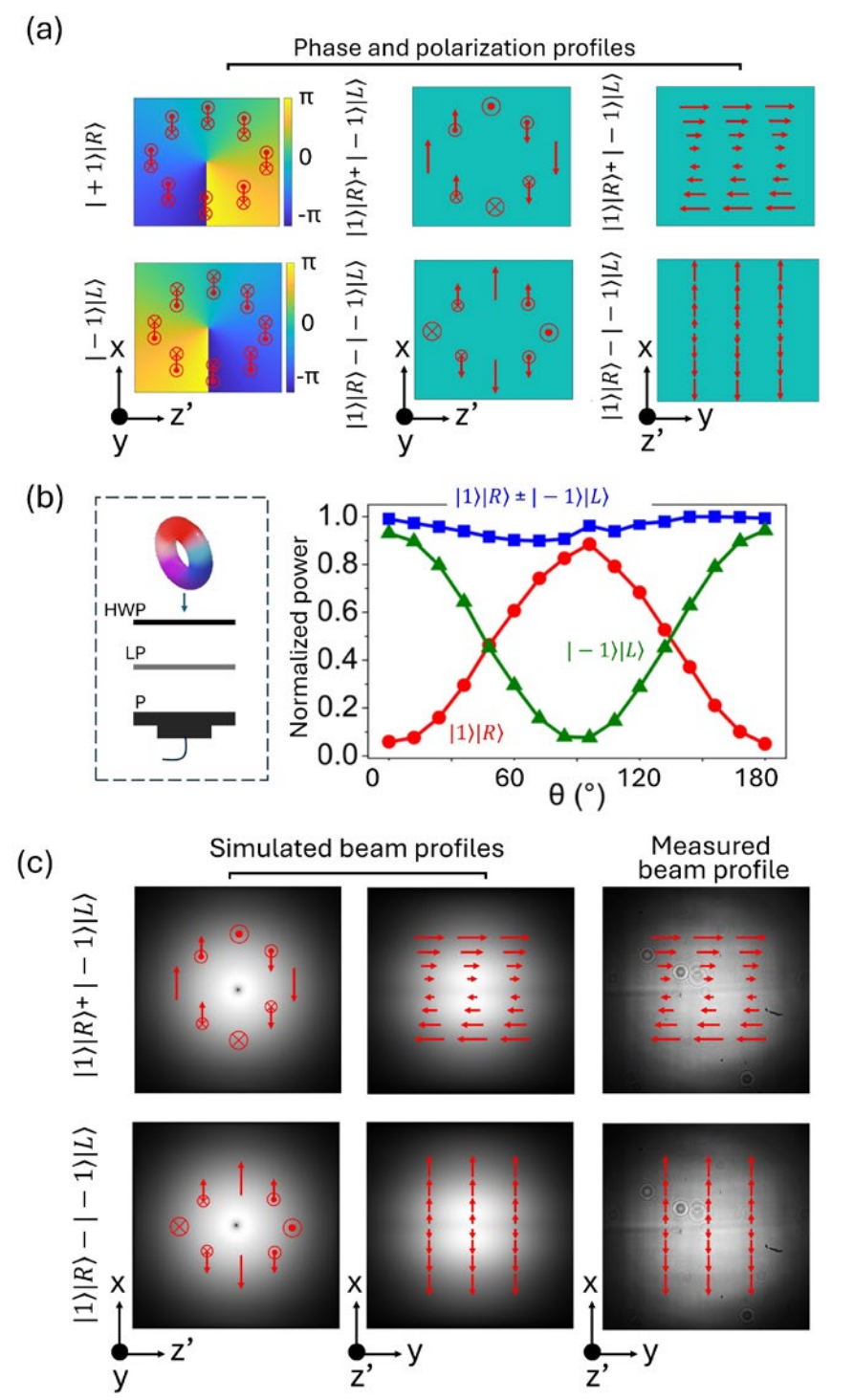

**Figure 3.** Polarization analysis results of t-OAM and circular polarization entangled spatiotemporal structured light. (a) Illustration of simulated phase (color map) and polarization (red arrows) profiles associated with various states of light in the x-z' plane. For mode entangled light, the polarization becomes spatiotemporal dependent, a new type of vector beam. (b) Polarization analyzer results with input states of light as either |1⟩|R⟩ (red curve), |-1⟩|L⟩ (green curve), or |1⟩|R⟩±|-1⟩|L⟩ (blue curve). Their respective visibilities are V = 0.89, 0.85, and 0.05. The schematic of the polarization analyzer is shown on the left. HWP: half-wave plate; LP: linear polarizer; P: power meter. (c) Simulation of the intensity and polarization profiles of mode entangled light in the x-z' plane (left) and x-y plane (middle). The corresponding experimentally measured intensity profiles in the x-y plane are shown on the rightmost column. Red arrows illustrate simulated local polarization directions.

respectively. In each of the panels, the first column depicts the simulated spatiotemporal phase profiles of the in-phase and out-of-phase superpositions of the two beams. The second and third columns show the corresponding simulated intensity profiles at the x-z' and x-y planes, respectively, where $z' = z - v_g t$. The dotted red lines between the second and third columns highlight the fact that the latter is a time-projection of the former. The last column shows experimentally measured beam intensity profiles, obtained by projecting 3D pulsed beams onto the x-y plane and captured with a 2D beam profiler. Fig. 2(b) shows constructive and destructive interference between two regular Gaussian beams, both expressed as |q=0⟩ and non-orthogonal to each other. As expected,

uniformly bright and dark intensity profiles were observed for in- and out-of-phase conditions, respectively. This straightforward experiment is meant to serve as a control for subsequent experiments involving t-OAM states.

Next, the top row of Fig. 2(c) shows the in-phase superposition of two beams carrying two different t-OAM charges, |q=0⟩ and |q=1⟩, resulting in constructive and destructive interference at the top and bottom half of the beam, respectively. In contrast, when the two beams were out-of-phase, the areas where constructive and destructive interference flipped. This phenomenon occurs because where the phase dislocations occur dictates where there would be partial destructive interferences and, thus, vanished intensities at those regions. For in-phase superposition of |q=0⟩ and |q=1⟩, the π-phase dislocation occurs at the six o'clock position. However, for out-of-phase conditions, the π-phase dislocation flipped to the twelve o'clock position. As such, the off-centered half-beam intensity profiles seen on the x-y plane are characteristic of the superimposition between t-OAM states of light with unequal magnitude of topological charges |q|. Conversely, when the two beams have the same charge q but opposite signs, the intensity profiles in both x-y and x-z' planes are symmetric about the middle and center lines, regardless of whether they were in- or out-of-phase. For instance, when two beams represented by |q=+1⟩ and |q=-1⟩ are in-phase, we will observe a symmetric profile composed of two bright lobes separated by a horizontal zero-intensity line (top row of Fig. 2(d)). The location of low intensities, which indicates partial destructive interferences, coincide with the horizontal π-phase jump dislocation line on the $k_x$-ω plane of the pulse shaper occurs. In comparison, the out-of-phase condition results in a vertical phase dislocation through the middle, which led to a corresponding region of low brightness in the intensity profile in the x-z' plane. However, as expected, the bi-lobed appearance cannot be observed after projection along z'. Instead, a horizontally elongated lobe located at the center of the intensity profile (bottom row of Fig. 2(d)) was obtained. Our experimental observations (rightmost column of panels fig. 2(b)-(d)) agree well with our simulations (third column of panels fig. 2(b)-(d)).

For orthogonality tests, we leverage the fact that orthogonal states of light do not interfere with each other and thus their superposition states would exhibit constant power as their relative phase varies. This equates to zero visibility, based on the definition of visibility $V=|P_{max}-P_{min}|/(P_{max}+P_{min})$. By using six example combinations of different topological charges as shown in Fig. 2(b) – (d), we experimentally established the orthogonality between the family of various t-OAM states. In these tests, we acquired time-averaged power of the superimposed pulsed beam during in-phase and out-of-phase conditions to determine their interferability through visibility measurements. In the case of a pair of conventional Gaussian pulsed beams, which is equivalent to |q=0⟩, the measured beam intensity profiles exhibited a maximum total power of 2.7 mW during in-phase superposition and constructive interference (Fig.2(b), top). The average power dropped to a minimum of 0.9 mW during out-of-phase superposition and destructive interference (Fig.2(b), bottom). The visibility calculated from these two conditions is V=0.5. In contrast to the superposition of Gaussian beams, the visibility of the superposition between |q=0⟩ and |q=1⟩ is near zero at V=0. 012, thus confirming their orthogonality. The orthogonality between |q=1⟩ and |q=-1⟩ were similarly verified when a near

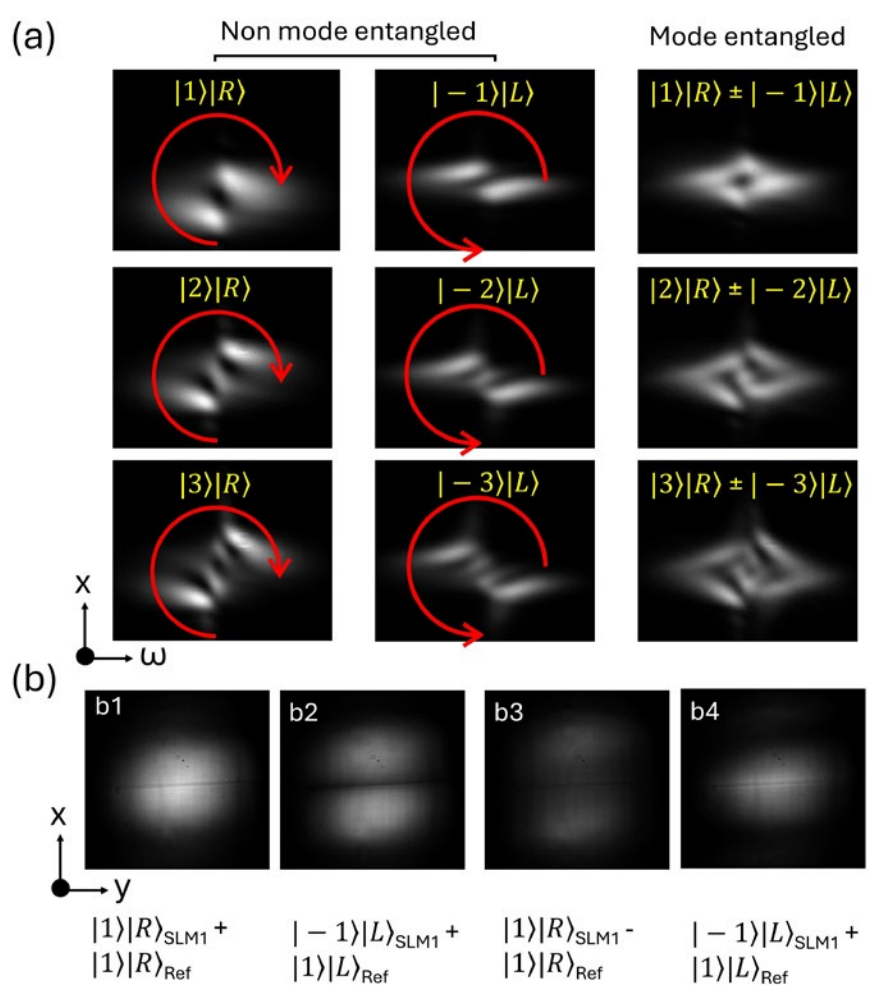

**Figure 4.** Experimental characterization of mode entangled beams. (a) Imaging spectrometer-based characterization on the x-ω plane. Left: Experimentally measured spatial-spectral intensity profiles of right-hand circularly polarized beams carrying t-OAM charges 1,2, and 3. Middle: Experimentally measured spatial-spectral intensity profiles of left-hand circularly polarized beams carrying t-OAM charges -1, -2, and -3. Right: Experimentally measured spatial-spectral intensity profiles of mode entangled beams with different t-OAM charges, where each of the three profiles have two possibles, in-phase or out-of-phase, superpositions. Red arrows indicate polarization. (b) Direct intensity projection-based characterization on the x-y plane. The images were used to resolving the phase ambiguity (in-phase or out-of-phase mode entangled light) present in the data captured by the imaging spectrometer. Top: Projected intensities of $(|1\rangle_{SLM1}\pm|1\rangle_{Ref})|R\rangle$ where subscripts SLM1 and Ref represent beams from the first SLM and the third SLM (which is used for generating a reference beam), respectively. Bottom: Projected intensities of $(|-1\rangle_{SLM2}\pm|1\rangle_{Ref})|L\rangle$.

zero visibility of V=0.021 were measured. Experimental results demonstrating the orthogonality of additional combinations of identical and different spatiotemporal topological charges can be found in Supplementary Materials Fig. S1. From this series of experiments, we conclude that t-OAM with different topological charges is indeed orthogonal to one another.

**4. Mode entanglement with t-OAM and polarization.** We next present results of mode entanglement with t-OAM and circular polarization. A representative case of this new family of mode entangled light can be written as |Ψ⟩=|+q⟩|R⟩±|-q⟩|L⟩. The experimental setup shown in Fig. S2 of Supplementary Materials was used to generate and analyze this family of mode entangled light. Briefly, this setup comprises five subsystems: the mode entanglement subsystem, the reference subsystem, the polarization analyzer, the imaging spectrometer, and the beam profiler. In the modified 4f pulse shaper, the laser beam was split into a mode entanglement subsystem and a reference subsystem. In the mode entanglement subsystem, a half-wave plate (HWP1 in Fig. S2) set at 22.5° is used to turn the polarization vector of the initial s-polarized beam 45° diagonally. This diagonally polarized light (denoted as D) is then sent to a polarizing beam splitter (PBS1), which enables equal powered s- and p-polarized light to be directed towards SLM1 and SLM2, respectively, for structuring. The combination of the return light from these two SLMs would result in the entanglement between t-OAM and linear polarizations, such as diagonal (D) and anti-diagonal (A) polarization. This state of the light can be expressed as |Ψ⟩=|+1⟩|A⟩±|-1⟩|D⟩. Subsequently, a quarter-wave plate (QWP1 in Fig. S2) was used to convert that into t-OAM and circular polarization mode entangled light, which can be expressed as |Ψ⟩=|+1⟩|R⟩±|-1⟩|L⟩. The reference subsystem is an essential tool that allows us to distinguish in-phase and out-of-phase conditions required for generating mode entangled beams. Therefore, the mode entanglement and reference sub-systems were used in tandem during experiments. A detailed discussion of this modified setup can be found in the Supplementary Materials.

Given the simplest case, |Ψ⟩=|+1⟩|R⟩±|-1⟩|L⟩, figure 3(a) shows simulated phase profiles (color map) and local polarization directions (red arrows) associated with various states of light, including |Ψ⟩=|+1⟩|R⟩ (top left), |Ψ⟩=|-1⟩|L⟩ (bottom left), and |Ψ⟩=|+1⟩|R⟩±|-1⟩|L⟩ (middle column). When |+1⟩|R⟩ and |-1⟩|L⟩ were coherently combined, we no longer have a simple scalar field. Instead, a space- and time-dependent vector beam is generated, exhibiting inhomogeneous electric field directions at different points in space and time. Furthermore, the local field directions vary with different spatiotemporal phase, $\Phi_{ST}$(x, z'), across the x-z' plane. We show how spatiotemporal dependent polarization arise in the derivations presented in Section 3 of the Supplementary Material. The result is illustrated in Fig. 3(a), with the red arrows indicating the local polarization vectors. As shown in the middle column of Fig. 3a, the mode entangled pulse |+1⟩|R⟩±|-1⟩|L⟩ possesses vertical (along the x-axis) and out-of-plane (along the y-axis) linear polarizations that varied in strength from the beam center to edge of the x-z' plane. When projected onto the x-y plane, the time-averaged pulse exhibits either only horizontal or vertical polarization, the strength of which is x-dependent.

As a method to verify the polarization states of the mode entangled light, we directed the light exiting the pulse shaper to a polarization analyzer (left of Fig. 3(b)). To facilitate the analysis of the polarization states of the generated light, another quarter-wave plate (QWP2) was used to convert the circular polarization states into linear polarization states, without compromising the validity of our analysis. The optical power that reached the power meter was recorded as a function of the angle of the half-wave plate in the polarization analyzer. The normalized power as a function of angle, θ, between the polarization vector of the incident beam and the linear polarizer, are shown in a graph in Fig. 3(b). Our experimental results demonstrate that |1⟩|R⟩ (red curve) and |-1⟩|L⟩ (green curve) exhibit sinusoidal variations in the measured power as a function of θ, with the maximum of the former aligning with the minimum of the latter. On the other hand, the mode entangled vector beam, |+1⟩|R⟩±|-1⟩|L⟩ (blue curve), is insensitive to the angle θ. This is because for the given spatiotemporal dependent polarization profile (middle column of Fig. 3(a)), its time-averaged and position-averaged transmission through the analyzer is constant regardless of the rotation angle of the HWP. The respective visibilities corresponding to|1⟩|R⟩, |-1⟩|L⟩, and |+1⟩|R⟩±|-1⟩|L⟩ experimentally measured to be V=0.89, 0.85, and 0.05, respectively. Through the sinusoidal variation of the measured power with θ in the polarization analyzer, the visibilities measurements, and the coincidence of the minimum of the curve for |-1⟩|L⟩ with the maximum of the curve for |1⟩|R⟩, we verified that the intended optical states were successfully generated.

In addition to the simulated phase profiles, we also show simulated intensity profiles of mode entangled light in the x-z' plane (left) and x-y plane (middle) in Fig. 3(c), with polarization directions illustrated by red arrows. The intensity profiles of both |1⟩|R⟩+|-1⟩|L⟩ and |1⟩|R⟩-|-1⟩|L⟩ has a small dark spot in the middle due to phase singularity. As expected, both the simulated intensity profile in the x-y plane show decreased intensity across the middle of the profile, which arose from the z'-projection of a ring-like intensity pattern x-z' plane. These simulated intensity profiles on the x-y plane serve as a comparison to the experimentally measured intensity profile captured with a beam profiler without the use of time-resolved imaging techniques, as shown in the rightmost column of Fig. 3(c). The schematic of the beam profiler is shown in Fig. S2 for this measurement. Our experimental results shown on the rightmost column of Fig. 3(c) agree well with the simulations. Additionally, the lack of interference between these two intensity profile projections confirms the orthogonality of the polarization states $|R\rangle$ and $|L\rangle$ in the mode entangled light generated by our system.

We further characterized the mode entangled light, |Ψ⟩=|+q⟩|R⟩±|-q⟩|L⟩, with different spatiotemporal topological charges q. Here we adapted the imaging spectrometer method that was invented in our prior work [18]. In this method, a spatially resolved spectral interferometer becomes an imaging spectrometer when used without a reference beam. This method is based on measuring spatial-spectral (x-ω) or wavevector-spectral ($k_x$-ω) profiles of a spatiotemporal structured beam carrying t-OAM by Fourier transform of t-OAM along the x-axis and the z'-axis, either simultaneously or in sequence. The main characteristics of t-OAM, such as its topological charge q value and helicity (sign; namely, the winding direction of spatiotemporal phase), can be identified directly from the unique and unambiguous profile features directly seen on the raw data from a beam profiler. Charge q can be determined by the numbers of dark regions in the multi-lobed structure along the diagonal directions. That is, one dark region indicates |q|=1. The sign of the charge (namely, the helicity) can be distinguished based on either diagonal or anti-diagonal spread of the lobes. See Fig. S2 for the detailed experimental setup of our imaging spectrometer implementation.

Figure 4(a) shows our experimentally measured spatial-spectral profiles of |1⟩|R⟩, |2⟩|R⟩, and |3⟩|R⟩ from SLM1, |-1⟩|L⟩, |-2⟩|L⟩, and |-3⟩|L⟩ from SLM2, and |1⟩|R⟩±|-1⟩|L⟩ , |2⟩|R⟩±|-2⟩|L⟩, |3⟩|R⟩±|-3⟩|L⟩, from both SLM1 and SLM2, respectively. In left and middle columns of Fig. 4(a), it is evident that the number of dark regions correspond well to the charge |q| of the light being tested. Although they share the same number of dark regions, the direction which series of dark regions occur are different, with one along the diagonal and the other along the anti-diagonal direction. This is a visual sign that the light of possess opposite helicity (sign) of charge q. Our measurements of spatial-spectral profiles of mode

entangled light are shown in the right column of Fig. 4(a). Although we can still distinguish |q| from those entangled beams, we can no longer differentiate the helicity as both diagonal and anti-diagonal multi-lobed structures were superimposed onto each other. Additionally, the use of the imaging spectrometer alone does not allow in-phase and out-of-phase superpositions to be distinguished.

In address this ambiguity, we added a reference subsystem using a third SLM, SLM3 as shown in Fig. S2. In this subsystem, the beam is split into two paths. One of the arms carries s-polarization, while the other goes through a half-wave plate place at either 22.5° or -22.5°. This generates either diagonal or anti-diagonal polarization. The SLM3 in the reference subsystem is fixed with a LG(0,1) phase pattern. With reference to beamsplitter BS2, the optical path length of the light sent to the mode entangled subsystem is then either matched with that of the reference arm or set to have a $\pi$-phase difference, creating in-phase and out-of-phase conditions. We note that this subsystem is only used aligning the system to achieve the design in-phase or out-of-phase conditions. Once that is achieved, the reference beam is blocked in subsequent experiments. To demonstrate that this strategy allows us to resolve the phase ambiguity present in the imaging spectrometer data, we performed superpositions representing $(|1\rangle_{SLM1}+|1\rangle_{Ref})|R\rangle$ and $(|1\rangle_{SLM1}-|1\rangle_{Ref})|R\rangle$. Figure 4(b1) and 4(b3) show clear difference between the intensity images capture during in- and out-of-phase superpositions. Similarly, results corresponding to $(|-1\rangle_{SLM2}+|1\rangle_{Ref})|L\rangle$ and $(|-1\rangle_{SLM2}-|1\rangle_{Ref})|L\rangle$ show clear difference that enable the differentiation between in and out-of-phase conditions.

**5. Discussion and Conclusion.** In this study, we proposed a new family of classical entangled state of light, constructed by spatiotemporal structured vector light fields based on t-OAM. After verifying that t-OAM of different spatiotemporal topological charges are orthogonal states, we successfully leveraged t-OAM and polarization in the generation of a new form of mode entangled light. This new family of mode entangled light represents a new type of spatiotemporal couple vector beams that have non-separable, spatiotemporal-varying polarization and spatiotemporal phase profiles. The generation is implemented by a custom 4f pulse shaper design that includes a Michelson-like interferometer within it, enabling us to coherently combine two pulsed beams, one with t-OAM (e.g., topological charge q=+1, +2, …) and a polarization state (e.g., R, D, V), and the other with an opposite and orthogonal t-OAM (e.g., q=-1,-2, …) and an opposite polarization state (e.g., L, A, H).

We have also developed and used several methods to characterize our mode entangled light. This includes the use of a polarization analyzer, an imaging spectrometer, and introducing a reference beam within the pulse shaper to address phase ambiguity. These methods allow us to confirm polarization states, t-OAM charges, and in-phase or out-of-phase superposition of mode entanglement. We note that a new spatiotemporal resolved polarimetry method is needed for full characterization of the polarization pattern across the mode entangled light we generated, but there are currently no suitable techniques capable of accomplishing that. As a matter of fact, a full characterization for any kind of spatiotemporal structured light fields—including all DoFs in coupled, non-separable space, time, spectrum, phase, polarization states, is much needed for the whole community.

In conclusion, our successful demonstration of utilizing t-OAM for mode entanglement open new opportunities for applications in quantum-inspired sensing, imaging, and higher dimensional flying qubit. This new development and robust framework hold significant promise for advancing quantum and classical applications alike in the future based on spatiotemporal structured light fields.

**Acknowledgments** The authors acknowledge Kaidi Fan from the Department of Informatics, Indiana University, Bloomington, for preparing scientific illustrations and schematics used in our graphical abstracts and the first figure.

**Research funding:** The authors acknowledge support from U.S. Department of Energy (DOE), Office of Biological and Environmental Research (BER), grant no. DE-SC0023314 and DE-SC0025194.

**Author contribution:** CTL and HML co-direct the project. HCH, HML, and CTL designed the experiments. HCH and KM conducted experiments. All authors analyzed the data and prepared the manuscript. All authors have accepted responsibility for the entire content of this manuscript and approved its submission.

**Conflict of interest**: Authors state no conflict of interest.

**Data availability statement**: The data reported in this work are available from the corresponding authors upon reasonable requests.

## References

[1] S. Slussarenko and G. J. Pryde, "Photonic quantum information processing: A concise review." Appl. Phys. Rev. vol. 6, p. 041303, 2019. https://doi.org/10.1063/1.5115814.

[2] Z. Zhang et al., “Entanglement-based quantum information technology: a tutorial,” Adv Opt Photonics, vol. 16, no. 1, p. 60, Mar. 2024, doi: 10.1364/aop.497143.

[3] R. J. C. Spreeuw, “A Classical Analogy of Entanglement,” Found Phys, vol. 28, no. 3, pp. 361–374, 1998, doi: 10.1023/A:1018703709245.

[4] A. Luis, “Coherence, polarization, and entanglement for classical light fields,” Opt Commun, vol. 282, no. 18, pp. 3665–3670, 2009, doi: https://doi.org/10.1016/j.optcom.2009.06.024.

[5] A. Aiello, F. Tö Ppel, C. Marquardt, E. Giacobino, and G. Leuchs, “Quantum-like nonseparable structures in optical beams,” New J Phys, vol. 17, Apr. 2015, doi: 10.1088/1367-2630/17/4/043024.

[6] M. Hillery and M. S. Zubairy, “Entanglement conditions for two-mode states,” Phys Rev Lett, vol. 96, no. 5, 2006, doi: 10.1103/PhysRevLett.96.050503.

[7] S. Lu et al., “Operator analysis of contextuality-witness measurements for multimode entangled single-neutron interferometry,” Phys Rev A (Coll Park), vol. 101, no. 4, Apr. 2020, doi: 10.1103/PhysRevA.101.042318.

[8] X.-F. Qian, B. Little, J. C. Howell, and J. H. Eberly, “Shifting the quantum-classical boundary: theory and experiment for statistically classical optical fields,” Optica, vol. 2, no. 7, p. 611, Jul. 2015, doi: 10.1364/optica.2.000611.

[9] L. Allen, M. W. Beijersbergen, R. J. C. Spreeuw, and J. P. Woerdman, “Orbital angular momentum of light and the transformation of Laguerre-Gaussian laser modes,” Phys Rev A (Coll Park), vol. 45, no. 11, pp. 8185–8189, Jun. 1992, doi: 10.1103/PhysRevA.45.8185.

[10] K. Y. Bliokh and F. Nori, “Spatiotemporal vortex beams and angular momentum,” Phys Rev A (Coll Park), vol. 86, no. 3, p. 33824, Sep. 2012, doi: 10.1103/PhysRevA.86.033824.

[11] K. Y. Bliokh and F. Nori, “Transverse and longitudinal angular momenta of light,” Phys Rep, vol. 592, pp. 1–38, 2015, doi: https://doi.org/10.1016/j.physrep.2015.06.003.

[12] A. P. Sukhorukov and V. V. Yangirova "Spatio-temporal vortices: properties, generation and recording", Proc. SPIE 5949, Nonlinear Optics Applications, 594906 (5 October 2005); https://doi.org/10.1117/12.623906.

[13] N. Jhajj, I. Larkin, E. W. Rosenthal, S. Zahedpour, J. K. Wahlstrand, and H. M. Milchberg, “Spatiotemporal optical vortices,” Phys Rev X, vol. 6, no. 3, 2016, doi: 10.1103/PhysRevX.6.031037.

[14] S. W. Hancock, S. Zahedpour, A. Goffin, and H. M. Milchberg, “Free-space propagation of spatiotemporal optical vortices,” Optica, vol. 6, no. 12, p. 1547, Dec. 2019, doi: 10.1364/optica.6.001547.

[15] A. Chong, C. Wan, J. Chen, and Q. Zhan, “Generation of spatiotemporal optical vortices with controllable transverse orbital angular momentum,” Nat Photonics, vol. 14, no. 6, pp. 350–354, Jun. 2020, doi: 10.1038/s41566-020-0587-z.

[16] S. Huang, P. Wang, X. Shen, J. Liu, and R. Li, “Diffraction properties of light with transverse orbital angular momentum,” Optica, vol. 9, no. 5, p. 469, May 2022, doi: 10.1364/optica.449108.

[17] S. Zahedpour, S. W. Hancock, and H. M. Milchberg, “Transient grating single-shot supercontinuum spectral interferometry (TG-SSSI),” in Optics InfoBase Conference Papers, Optica Publishing Group (formerly OSA), Nov. 2020. doi: 10.1364/ol.417803.

[18] G. Gui, N. J. Brooks, B. Wang, H. C. Kapteyn, M. M. Murnane, and C. T. Liao, “Single-Frame Characterization of Ultrafast Pulses with Spatiotemporal Orbital Angular Momentum,” ACS Photonics, vol. 9, no. 8, pp. 2802–2808, Aug. 2022, doi: 10.1021/acsphotonics.2c00626.

[19] S. Zahedpour, S. W. Hancock, and H. M. Milchberg, “Second harmonic generation of spatiotemporal optical vortices (STOVs) and conservation of orbital angular momentum,” in Optics InfoBase Conference Papers, Optica Publishing Group (formerly OSA), 2021. doi: 10.1364/optica.422743.

[20] G. Gui, N. J. Brooks, H. C. Kapteyn, M. M. Murnane, and C. T. Liao, “Second-harmonic generation and the conservation of spatiotemporal orbital angular momentum of light,” Nat Photonics, vol. 15, no. 8, pp. 608–613, Aug. 2021, doi: 10.1038/s41566-021-00841-8.

[21] A. B. Stilgoe, T. A. Nieminen, and H. Rubinsztein-Dunlop, “Controlled transfer of transverse orbital angular momentum to optically trapped birefringent microparticles,” Nat Photonics, vol. 16, no. 5, pp. 346–351, May 2022, doi: 10.1038/s41566-022-00983-3.

[22] J. Huang, J. Zhang, T. Zhu, and Z. Ruan, “Spatiotemporal Differentiators Generating Optical Vortices with Transverse Orbital Angular Momentum and Detecting Sharp Change of Pulse Envelope,” Laser Photon Rev, vol. 16, no. 5, May 2022, doi: 10.1002/lpor.202100357.

[23] S. Huang et al., “Spatiotemporal vortex strings,” Sci Adv, vol. 10, no. 19, p. eadn6206, Dec. 2024, doi: 10.1126/sciadv.adn6206.

[24] M. Erhard, M. Krenn, and A. Zeilinger, “Advances in high-dimensional quantum entanglement,” Nature Reviews Physics, vol. 2, no. 7, pp. 365–381, 2020, doi: 10.1038/s42254-020-0193-5.

[25] W. Chen, W. Zhang, Y. Liu, F. C. Meng, J. M. Dudley, and Y. Q. Lu, “Time diffraction-free transverse orbital angular momentum beams,” Nat Commun, vol. 13, no. 1, Dec. 2022, doi: 10.1038/s41467-022-31623-7.

Supplemental Material for
# Transverse orbital angular momentum and polarization entangled spatiotemporal structured light

Hsiao-Chih Huang[1,2], Kefu Mu[2], Hui Min Leung[2*], and Chen-Ting Liao[1*]
[1]Department of Physics, Indiana University, Bloomington, Indiana 47405, U.S.A.
[2]Department of Intelligent Systems Engineering, Indiana University, Bloomington, Indiana 47405, U.S.A

## Section 1. Theory of the superpositions of t-OAM modes

The ideal form of the wave function of line-STOVs with a spatiotemporal topological charge $q$ can be expressed as a function in cylindrical coordinates, $\psi(\rho,\Phi)$, with the radial coordinate, $\rho(x,t)=\rho(x,z')=\sqrt{x^2+(z')^2}$, and azimuthal coordinate, $\Phi(\mathrm{x},\mathrm{t})=\Phi(\mathrm{x},\mathrm{z}')=tan^{-1}(x/z')$. Here, $z'=z-v_g t$ represents the space-like coordinate in the moving reference frame of a pulsed beam travelling with group velocity $v_g$. The wave function of t-OAM of light reads

$$\psi(\rho,\Phi)=\langle\rho,\Phi|\psi(\rho,\Phi)\rangle=f(\rho)e^{iq\Phi}, \quad (1)$$

where $|\psi(\rho,\Phi)\rangle$ is the quantum state of t-OAM of a photon, denotated by Dirac's *ket* notation and $f(\rho)$ is a function of the beam's parameters. The expectation value of energy of our t-OAM of light can be formulated within the framework of quantum mechanics using the photon energy operator $\hat{H}$ as:

$$\langle E\rangle=\frac{\int_0^a\int_0^{2\pi}\hat{H}|\psi(\rho,\Phi)|^2d\Phi d\rho}{\int_0^a d\rho\int_0^{2\pi}d\Phi}=\frac{1}{2\pi a}\int_0^a\int_0^{2\pi}\hat{H}\,|\psi(\rho,\Phi)|^2d\Phi d\rho, \quad (2)$$

where $a$ is the beam radius and the operator is defined as $\hat{H}=n\hbar\omega\hat{I}$, where n is the photon number (n = 0, 1, 2, …), $\hat{I}$ is the identity operator, and ω is the central frequency of the light beam. Therefore, we can rewrite the expectation value of energy as:

$$\langle E\rangle=\frac{1}{2\pi a}\int_0^a\int_0^{2\pi}n\hbar\omega\hat{I}\ |\psi(\rho,\Phi)|^2d\Phi d\rho=n\hbar\omega A, \quad (3)$$

where $A=\frac{1}{a}\int_0^a|f(\rho)|^2\,d\rho$ is a constant derived from the radial variable related to the light beam used. Now we consider averaged power from our pulsed light beam, $\langle P\rangle=\langle E\rangle/T=n\hbar\omega A/T$, where $T$ is our nominal data acqusition time (averaging time) and $T\gg t_0\gg\frac{2\pi}{\omega}$, $t_0$ is the nominal pulse duration (~200 fs) and $\frac{2\pi}{\omega}$ is one optical cycle (~3.4 fs) . When taking n = 1, we get the averaged, single-photon level expression of our t-OAM of light.

A light beam created by the superposition of two t-OAM light ($|\psi_1\rangle$ and $|\psi_2\rangle$) with topological charges $q_1$ and $q_2$, respectively, with relative phase, $\delta$, is expressed in a normalized quantum state as:

$$|\psi(\rho,\Phi)\rangle=\frac{|\psi_1\rangle+e^{i\delta}|\psi_2\rangle}{\sqrt{\langle\psi_1|\psi_1\rangle+(e^{-i\delta}\langle\psi_2|)(e^{i\delta}|\psi_2\rangle)}}=\frac{1}{\sqrt{2}|f(\rho)|}\left(|\psi_1\rangle+e^{i\delta}|\psi_2\rangle\right). \quad (4)$$

The radial function $f(\rho)$ is assumed to be identical for those of the two t-OAM light, while the azimuthal function $e^{iq_1\Phi}$ and $e^{iq_2\Phi}$ correspond to the usage in Eq. (1). Substituting Eq. (4) into Eq. (2) gives:

$$\begin{aligned}\langle P\rangle &=\frac{1}{T}\frac{1}{2}\left(\frac{1}{2\pi a}\right)[\int_0^a d\rho\int_0^{2\pi}d\Phi\,\langle\psi_1|\rho,\Phi\rangle\langle\rho,\Phi|\hat{H}|\psi_1\rangle\\ &\quad+\langle\psi_2|\rho,\Phi\rangle\langle\rho,\Phi|\hat{H}|\psi_2\rangle\\ &\quad+e^{i\delta}\langle\psi_1|\rho,\Phi\rangle\langle\rho,\Phi|\hat{H}|\psi_2\rangle\\ &\quad+e^{-i\delta}\langle\psi_2|\rho,\Phi\rangle\langle\rho,\Phi|\hat{H}|\psi_1\rangle]\\ &=\frac{n\hbar\omega A}{2T}\left[1+1+\frac{e^{i\delta}}{2\pi}\int_0^{2\pi}e^{i(q_2-q_1)\Phi}\,d\Phi+\frac{e^{-i\delta}}{2\pi}\int_0^{2\pi}e^{-i(q_2-q_1)\Phi}\,d\Phi\right]. \end{aligned} \quad (5)$$

The two cross terms in the last two integrals of Eq. (5) evaluate to be 2π when $q_1=q_2$ and zero when $q_1\neq q_2$. For the superposition of two beams with identical t-OAM ($q_1=q_2$), the averaged power varies as

$$\langle P\rangle=n\hbar\omega A(1+cos\,\delta)/T. \quad (6)$$

Therefore, the maximum power is $P_{max} = 2n\hbar\omega A/T$, and the minimum power, $P_{min} = 0$, occur when the superposition is in-phase ($\delta = 0$) and out-of-phase ($\delta = \pi$), respectively. In contrast, for the superposition of different t-OAM ($q_1 \neq q_2$), the power remains constant, $P = n\hbar\omega A/T$, regardless of $\delta$.

Visibility, $V$, which is defined by the mean power integrated across the beam, is expressed as

$$V = |P_{max} - P_{min}|/(P_{max} + P_{min}). \quad (7)$$

Therefore, we conclude that when $q_1 = q_2$ and when $q_1 \neq q_2$, the visibility is 1 and 0, respectively. Equation (7) is what we used in our experiments and mentioned in the main manuscript.

**Section 2. Experimentally measured intensity profiles on the x-y plane for identical and orthogonal t-OAM.**

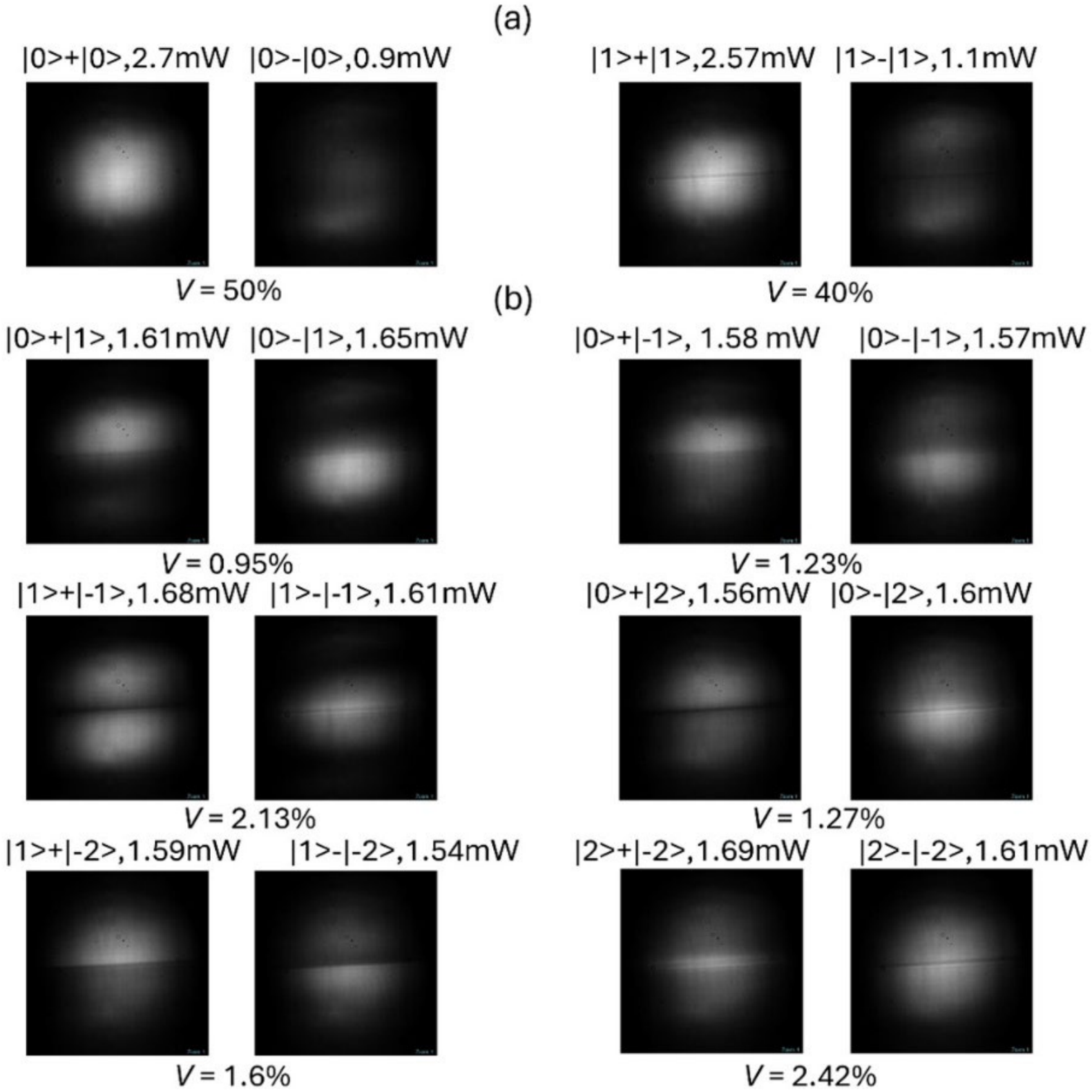

Figure S1. Experimentally measured intensity profiles on the x-y plane for the superpositions between (a) identical t-OAM with |0⟩±|0⟩ and |1⟩±|1⟩ and (b), different t-OAM with |0⟩±|1⟩, |0⟩±|-1⟩, |1⟩±|-1⟩, $|0\rangle \pm |2\rangle$, |1⟩±|-2⟩, and |2⟩±|-2⟩.

**Section 3. Schematic experimental setup for generating and analyzing mode entangled spatiotemporal structured light.**

The experimental setup for generating and analyzing mode entanglement is shown in Figure S2. This setup consists of a mode entanglement subsystem, a reference subsystem, a polarization analyzer, an imaging spectrometer, and a beam profiler.

Mode entanglement subsystem: The mode entanglement subsystem is similar to that depicted in Fig. 2(a), but with two key distinctions. Firstly, a thin plate polarizing beamsplitter (PBS1) replaces the thin plate nonpolarizing beamsplitter. Secondly, a half-wave plate (HWP1) angled at 22.5° is added to the system. These modifications enable the mode entanglement of t-OAM of $q_1$ = 1 and off-diagonal polarization (A1) with t-OAM of $q_2$ = -1 and diagonal polarization (D2), which can be expressed as |Ψ⟩ = |q=1⟩|A⟩ ± |q=-1⟩|D⟩. We note that, compared to the optical path toward SLM2, the path toward SLM1 includes an additional reflection from PBS1. Therefore, the phase pattern on SLM2 must be adjusted accordingly to compensate for the resulting OAM helicity change. The pair of orthogonal linear polarizations of A1 and D2 were subsequently converted to another pair of orthogonal circular polarizations, R1 and L2, after passing through a quarter-wave plate (QWP1) angled at 0°. At this point, we have generated our targeted mode entangled light, expressed as |Ψ⟩=|+1⟩|R⟩±|-1⟩|L⟩. This light is then directed towards three optical characterization subsystems, namely the polarization analyzer, imaging spectrometer, and beam profiler.

Polarization analyzer: The polarization analyzer was used to verify the polarization states generated by the mode entanglement subsystem described above. To aid the polarization analysis, the orthogonal pair of circular polarizations were first converted into linear polarizations (S1 and P2) through the use of a quarter-wave plate (QWP2) angled as 45°. The light was then sent through a rotatable half-wave plate and linear polarizer (LP) before the time-averaged optical power of the entire beam was captured by a power meter (P).

Imaging spectrometer: The t-OAM charges |+1⟩ and |-1⟩ can be distinguished by an imaging spectrometer as follows. First, two beams pass through a cylindrical lens (CL2v) with a focal length of f=70 cm. The beams are diffracted by a thin plate transmission grating (G2) with a groove density of 1600 lines/mm and then focused by a vertical cylindrical lens (CL3v) with a focal length of f=13 cm. The resultant profiles are captured on a beam profiler camera (C2), which is located at the focal plane of CL2v.

Reference subsystem: It is crucial to introduce a phase shift of either 0 (namely, $e^{i\delta} = e^{i(0)} = e^{i(2\pi)} = +1$) or π (namely, $e^{i\delta} = e^{i(\pi)} = -1$ ) between |1⟩|R⟩ and |-1⟩|L⟩ to achieve in-phase or out-of-phase superpositions, respectively. Because the imaging spectrometer does not allow these two conditions to be distinguished, a reference subsystem described here was used. The SLM3 in this subsystem is loaded with a phase pattern of LG(0,1) to generate a beam with |1⟩. The first index in the LG pattern is set to zero as we only used the azimuthal index for OAM. When the original s-polarized light is sent into the reference subsystem, it is split into two arms by a beamsplitter (BS3). In one arm, the light is converted into polarization of either A3 or D3 by using a half-wave plate (HWP2) oriented at ±22.5°. After exiting the reference subsystem, these states were subsequently converted to circular polarizations (i.e., R3 or L3) by passing through QWP1 at 0°. This reference beam is then overlapped with the light returning from either SLM1 (i.e., |1⟩|R⟩) or SLM2 (i.e., |-1⟩|L⟩) individually. We achieve in-phase superposition to generate |+1⟩|R⟩+|-1⟩|L⟩ by adjusting the translational stages controlling SLM1 and SLM2 to ensure constructive interference between both the reference and |1⟩|R⟩, as well as the reference and |-1⟩|L⟩. Conversely, the out-of-phase superposition is achieved by introducing an additional $\pi$ phase delay in either arm associated with |1⟩|R⟩ or |-1⟩|L⟩. This is again accomplished by adjusting the translational stages of either SLM1 or SLM2 to create constructive interference between the reference and |1⟩|R⟩ and destructive interference between the reference and |-1⟩|L⟩.

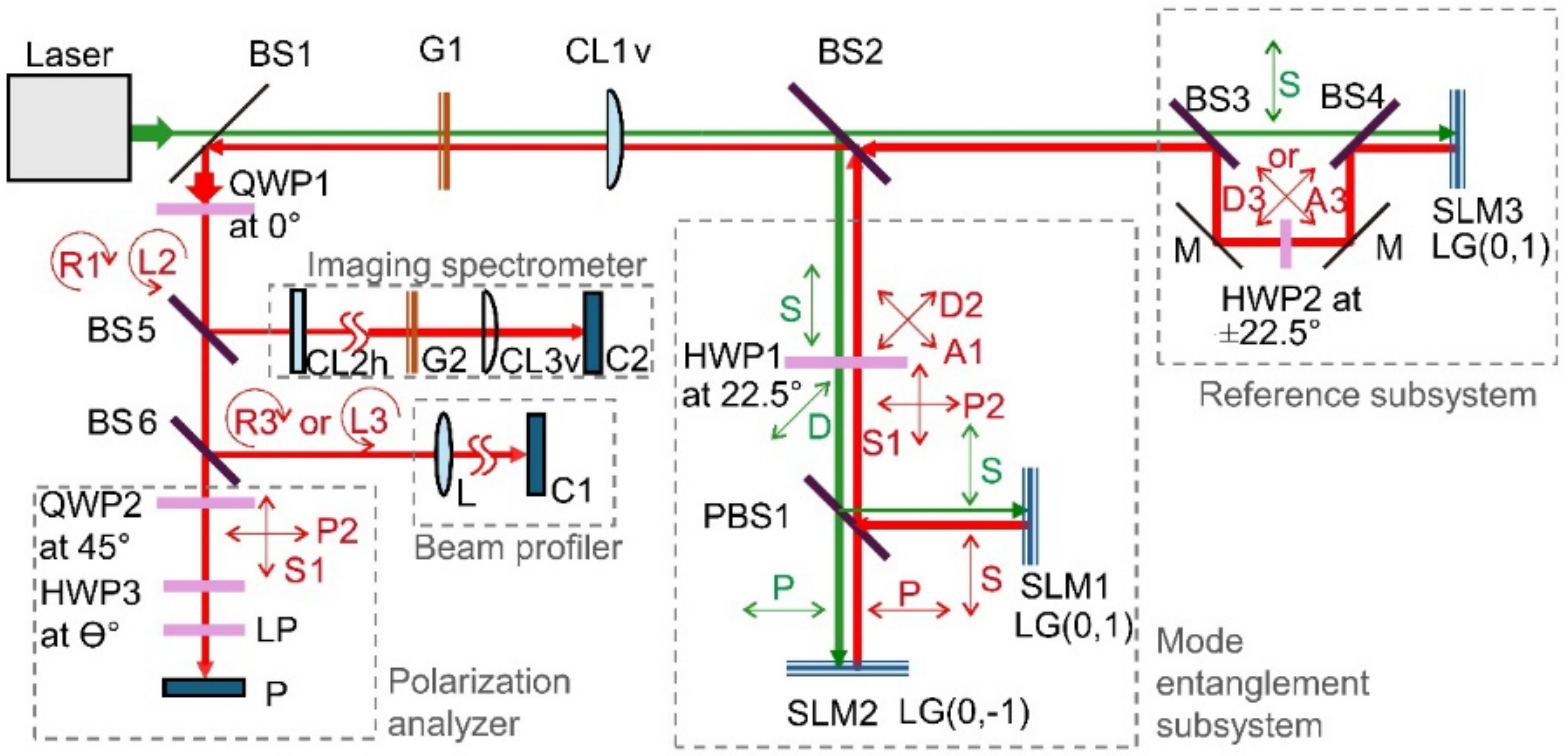

Figure S2. Experimental setup for generating and analyzing the mode entanglement of t-OAM and polarization. This comprises a folded 4*f* pulse shaper and a Michelson interferometer with HWP1 and PBS1, a reference subsystem, an imaging spectrometer, a beam profiler, and a polarization analyzer, indicated by five dashed boxes. BS: thin plate non-polarizing beam splitter; PBS: thin plate polarizing beam splitter; HWP: half-wave plates; QWP: quarter-wave plates; G: transmission grating; CLh and CLv: cylindrical lenses with horizontal and vertical focal lines; L: spherical lens; SLM: spatial light modulator; C: camera; P: power meter.

## Section 4. Theory showing t-OAM and circular polarization mode entangled light are vector beams with spatiotemporal dependent polarization.

The electric field of a beam carrying t-OAM can be represented as

$$E(x,y,z,t) \propto Ue^{iq\Phi_{ST}(x,t)}e^{i(k_z z-\omega t)} = e^{iq\Phi_{ST}}e^{ik_z(z-v_g t)}$$

It propagates along the z-axis, where $z' = z - v_g t$ represents space-like coordinate in the moving reference frame of a laser pulse travelling with group velocity $v_g$, where $v_g \sim c$.

Neglecting the carrier term$e^{i(k_z z-\omega t)}$, the main feature of t-OAM stems from its spatiotemporal phase term that reads

$$e^{iq\Phi_{ST}(x,t)} = e^{iq\Phi_{ST}(x,z')} = [cos\ (q\Phi_{ST})] \pm i[\sin\ (\ q\Phi_{ST})], \text{ where } \Phi_{ST} = tan^{-1}(\frac{x/x_0}{z'/\eta z_0'}).$$

In the discussion below, we drop the subscript spatiotemporal and use $\Phi = \Phi_{ST}(x,z')$ for simplicity. We also assume $x_0 = z_0' = 1$ and $\eta = 1$ for a simplified case.

Right-hand circularly polarized (R) light and left-hand circularly polarized (L) light can be represented as

$$\begin{cases}|R(x,y)\rangle \\ |L(x,y)\rangle\end{cases} \propto |P(x,y)\rangle \pm i|S(x,y)\rangle \propto |H(x,y)\rangle \pm i|V(x,y)\rangle,$$

where $|P\rangle, |S\rangle, |V\rangle, |H\rangle$ are p-, s-, vertical-, horizontal- polarized light, respectively.

Here, $|P\rangle$=$|H\rangle$ is defined along the y-axis, and $|S\rangle$=$|V\rangle$ is defined along the x-axis.
Similarity, diagonal and anti-diagonal polarized light can be represented as

$$\begin{cases}|D(x,y)\rangle \\ |A(x,y)\rangle\end{cases} \propto |P(x,y)\rangle \pm |S(x,y)\rangle$$

The polarization depends on the position of beam profile in the x-z' plane for the in-phase and out-of-phase mode entangled light, $|\Psi(x,y,z')\rangle = |q=1\rangle|R(x,y)\rangle \pm |q=-1\rangle|L(x,y)\rangle$.

For a mode entangled light with first term and second term in-phase, we can write it as

$$|\Psi(x,y,z')\rangle = |1\rangle|R(x,y)\rangle + |-1\rangle|L(x,y)\rangle \\ \propto e^{iq\Phi(x,z')}|R(x,y)\rangle + e^{-iq\Phi(x,z')}|L(x,y)\rangle$$

Considering the simplest case when $q = 1$,

If $\Phi(x,z') = 0$ [twelve o'clock orientation at the x-z' plane], then we get $|\Psi(x,y,z')\rangle = |R(x,y)\rangle + |L(x,y)\rangle \propto |P(x,y)\rangle$

If $\Phi = \frac{\pi}{2}$ [three o'clock orientation at the x-z' plane], then we get $|\Psi(x,y,z')\rangle = i(|R(x,y)\rangle - |L(x,y)\rangle) \propto -|S(x,y)\rangle$

If $\Phi = \pi$, then we get $|\Psi\rangle = -(|R\rangle + |L\rangle) \propto -|P\rangle$

If $\Phi = \frac{3\pi}{2}$, we get $|\Psi\rangle = -i(|R\rangle - |L\rangle) \propto |S\rangle$

If $\Phi = \pm\frac{\pi}{4}$, then we get

$$|\Psi\rangle = e^{i\Phi}|R\rangle + e^{-i\Phi}|L\rangle \propto (1 \pm i)(|P\rangle + i|S\rangle) + (1 \mp i)(|P\rangle - i|S\rangle) \\ = [(|P\rangle \mp |S\rangle) \pm i(|P\rangle \pm |S\rangle)] + [(|P\rangle \mp |S\rangle) \mp i(|P\rangle \pm |S\rangle)] \\ \propto |P\rangle \mp |S\rangle \\ \propto \begin{cases} |A(x,y)\rangle \\ |D(x,y)\rangle \end{cases}$$

If $\Phi = \pm\frac{3\pi}{4}$, then we get

$$|\Psi\rangle = e^{i\Phi}|R\rangle + e^{-i\Phi}|L\rangle \propto (-1 \pm i)(|P\rangle + i|S\rangle) + (-1 \mp i)(|P\rangle - i|S\rangle) \\ = [-(|P\rangle \pm |S\rangle) \pm i(|P\rangle \mp |S\rangle)] + [-(|P\rangle \pm |S\rangle) \mp i(|P\rangle \mp |S\rangle)] \\ \propto -(|P\rangle \pm |S\rangle) \\ \propto \begin{cases} -|D(x,y)\rangle \\ -|A(x,y)\rangle \end{cases}$$

For a mode entangled light with first term and second term out-of-phase, we can write it as

$$|\Psi\rangle = |1\rangle|R\rangle - |-1\rangle|L\rangle \sim e^{i\Phi}|R\rangle - e^{-i\Phi}|L\rangle$$

If $\Phi = 0$, then we get $|\Psi\rangle = |R\rangle - |L\rangle \propto |S\rangle$

If $\Phi = \frac{\pi}{2}$, then we get $|\Psi\rangle = i(|R\rangle + |L\rangle) \propto |P\rangle$

If $\Phi = \pi$, then we get $|\Psi\rangle = -(|R\rangle - |L\rangle) \propto -|S\rangle$

If $\Phi = \frac{3\pi}{2}$, then we get $|\Psi\rangle = -i(|R\rangle + |L\rangle) \propto -|P\rangle$

If $\Phi = \pm\frac{\pi}{4}$, then we get

$$|\Psi\rangle = e^{i\Phi}|R\rangle - e^{-i\Phi}|L\rangle \propto (1 \pm i)(|P\rangle + i|S\rangle) - (1 \mp i)(|P\rangle - i|S\rangle) \\ = [(|P\rangle \mp |S\rangle) \pm i(|P\rangle \pm |S\rangle)] - [(|P\rangle \mp |S\rangle) \mp i(|P\rangle \pm |S\rangle)] \\ \propto |P\rangle \pm |S\rangle \propto \begin{cases} |D\rangle \\ |A\rangle \end{cases}$$

If $\Phi = \pm\frac{3\pi}{4}$, then we get

$$|\Psi\rangle = |R\rangle - e^{-i\Phi}|L\rangle \propto (-1 \pm i)(|P\rangle + i|S\rangle) - (-1 \mp i)(|P\rangle - i|S\rangle) \\ = [-(|P\rangle \pm |S\rangle) \pm i(|P\rangle \mp |S\rangle)] - [-(|P\rangle \pm |S\rangle) \mp i(|P\rangle \mp |S\rangle)] \propto -(|P\rangle \mp |S\rangle) \\ \propto \begin{cases} -|A\rangle \\ -|D\rangle \end{cases}$$

We show the local polarization directions in two planes of x-z' and x-y of two mode entangled spatiotemporal structured lights with $|\Psi\rangle=|+1\rangle|R\rangle\pm|-1\rangle|L\rangle$ in the Figs. 3(a) and 3(c) of the main manuscript. In the x-y plane, the time average polarizations for $|+1\rangle|R\rangle\pm|-1\rangle|L\rangle$ are all P (parallel to the y-axis) and S (parallel to the x-axis), respectively, and gradually increasing in magnitude along the x-axis from its beam center.